\documentclass[a4paper,10pt]{article}
\usepackage[english]{babel}
\usepackage[T1]{fontenc}
\usepackage[utf8]{inputenc}
\usepackage{amssymb}
\usepackage{physics}
\usepackage{amsmath}
\usepackage{graphicx}
\usepackage[unicode]{hyperref}
\usepackage{tikz}
\usepackage{color}
\usepackage{genyoungtabtikz}
\usepackage[title]{appendix}
\usepackage{indentfirst}
\usepackage{bm}
\usepackage{xhfill}
\usepackage{bbm}
\usepackage{multirow}
\usepackage{array}
\usepackage[centertableaux]{ytableau}
\usepackage{multicol}
\usepackage[shortlabels]{enumitem}
\usepackage{forest}
\usepackage{float}

\hypersetup{
    colorlinks=true,
    linkcolor=blue,
    filecolor=magenta,      
    urlcolor=cyan,
}
 
\usepackage{amsthm}

\theoremstyle{definition}

\usepackage[nottoc]{tocbibind}

\usepackage{authblk}

\title{\textbf{Shannon information entropy, soliton clusters and Bose-Einstein condensation in log gravity}}

\author[1,2]{\textbf{Yannick Mvondo-She}}
\affil[1]{National Institute of Theoretical and Computational Sciences,  School of Physics and Mandelstam Institute for Theoretical Physics,
University of the Witwatersrand, Johannesburg, Wits, 2050, South Africa}
\affil[2]{DSI-NRF Centre of Excellence in Mathematical and Statistical Sciences (CoE-MaSS), South Africa}
\affil[ ]{\texttt{vondosh7@gmail.com}}

\date{}

\begin{document}

\maketitle

\begin{abstract}
We give a probabilistic interpretation of the configurational partition function of the logarithmic sector of critical cosmological topologically massive gravity, in which the Hurwitz numbers considered in our previous works assume the role of probabilities in a distribution on cycles of permutations. In particular, it is shown that the permutations are distributed according to the Ewens sampling formula which plays a major role in the theory of partition structures and their applications to diffusive processes of fragmentation, and in random trees. This new probabilistic result together with the previously established evidence of solitons in the theory provide new insights on the instability originally observed in the theory. We argue that the unstable propagation of a seed soliton at single particle level induces the generation of fragments of defect soliton clusters with rooted tree configuration at multiparticle level, providing a disordered landscape. The Shannon information entropy of the probability distribution is then introduced as a measure of the evolution of the unstable soliton clusters generated. Finally, based on Feynman's path integral formalism on permutation symmetry in the $\lambda$-transition of liquid helium, we argue that the existence of permutation cycles in the configurational log partition function indicates the presence of Bose-Einstein condensates in log gravity.  
\end{abstract}

\tableofcontents

\section{Introduction}
Cosmological topologically massive gravity at the critical point (CCTMG) was first considered by Grumiller and Johansson a decade and a half ago \cite{Grumiller:2008qz}. Under a more relaxed class of boundary conditions than the standard Brown–Henneaux ones \cite{Brown:1986nw}, this three-dimensional gravity theory in anti-de Sitter (AdS) space-time with a negative cosmological constant and a gravitational Chern–Simons term features a new mode, the logarithmic primary which spoils the unitarity of the theory. Rather than rendering it inconsistent, the additional mode brings new and interesting perspectives into the theory. Firstly, within the AdS/CFT framework, the appearance of the log mode at the chiral point was used to provide a conjecture for a theory of gravity holographically dual to a logarithmic conformal field theory, a type of conformal field theory useful in describing various systems, among which two-dimensional turbulence, critical polymers, percolation, or systems with (quenched) disorder \cite{Grumiller:2013at}. Two important characteristics of LCFTs are that on one hand they exhibit a non-diagonalizable Hamiltonian due to the degeneracy of certain operators that together with their so called logarithmic partners decompose the operator spectrum into Jordan cells, and on the other hand they display logarithmic singularities in the correlation functions. The manifestation of these hallmarks of non-unitary conformal field theories in gravity theories at the chiral point lead to the conjecture of a more recent version of a class of dualities called the AdS$_3$/LCFT$_2$ correspondence, and to the coining of such gravity theories, log gravity \cite{Maloney:2009ck}. 

A significant result was obtained in the calculation of the one-loop partition function, and was found to agree with the partition function of an LCFT up to single particle \cite{Gaberdiel:2010xv}. Subsequently, the logarithmic contribution of the partition function (also called the log partition function) became a stimulating subject of study, and lead to several interesting outputs \cite{Mvondo-She:2018htn,Mvondo-She:2019vbx,Mvondo-She:2021joh,Mvondo-She:2022jnf}. 

The interpretation of the partition function from the point of view of solitons of integrable systems offers new explorable avenues in our attempt to better understand the log sector. On one hand, the reformulation of the log partition function as a $\tau$-function of the KP I integrable solitonic hierarchy describing solitons which are unstable with respect to transverse perturbations echoes the unstable aspect of CCTMG originally considered in \cite{Grumiller:2008qz}. Precisely, it brings to light the instability of solitonic structures in the theory, manifested by the appearance of a disordered ensemble of solitons. On the other hand, the log partition was also expressed in terms of the set of infinitely many symmetries of the Burgers equation, called the Burgers hierarchy. The Burgers equation is the simplest partial differential equation that combines nonlinear propagation effects as well as diffusive effects. Of considerable importance in the description of weak shock phenomena in gas dynamics, it can be linearized using a direct coordinate transformation to the diffusion equation. From the linearization, the infinite number of symmetries comes out of the Burgers equation leading to the definition of the Burgers hierarchy, also related to the $\tau$-function of the KP hierarchy \cite{kodama2017kp,kodama2018solitons}. The integrability properties of the aforementioned hierarchies displayed in the log partition function therefore suggests the instance of an unstable solitonic propagation of diffusive nature in the theory.  Described as a generating function of polynomials for random mappings from $\mathbb{CP}^1$ to $\mathbb{CP}^1$ related to integrable hierarchies, the log partition function was also defined as a multinomial expansion over rooted trees where subtrees cluster around a root according to a well constructed coalescent process. These results indicate a solitonic cluster structure of collective excitations in the theory. The evolution of the soliton clusters formed and the associated phenomena constitute the subject of this paper. 

In section 2, we motivate our probabilistic study of the log partition function by showing how it naturally arises from its generation of cycle structures of permutations. The relation between permutations and integer partitions is characterized by the Hurwitz numbers, which now appear as probabilities in a probability distribution on permutations and partitions. We point out a remarkable relationship between the Hurwitz numbers and the Ewens's sampling formula \cite{ewens1972sampling,crane2016ubiquitous}, well-known in population genetics as a one-parameter family of probability distributions on the set of all partitions of an integer $n$. More precisely, it is the probability distribution of the number of different types of genes and their frequencies under neutral evolution, $i.e$ in the absence of selection. Also known among probabilists and statisticians as the multivariate Ewens distribution, the Ewens sampling formula finds applications in a very broad range of areas of physical and mathematical sciences such as in the calculation of multiparticle Veneziano amplitudes \cite{kholodenko2008landau}, combinatorial stochastic
processes \cite{kingman1978random,pitman1995exchangeable}, Macdonald polynomials \cite{diaconis2012probabilistic}, algebraic number theory \cite{billingsley1972distribution,donnelly1993asymptotic}. and even the Faà di Bruno’s formula for the derivative of a composite function \cite{hoppe2008faa}, whose Bell polynomial avatar became the premise of our work on the log partition function. A characteristic of the Ewens sampling formula is its description of theories in which dynamics are diffusive in nature. The evolution of the diffusive dynamics captured by the log partition function encoding the elegant Ewens distribution concurs with the representation of the partition function in terms of the Burgers hierarchy, and goes in the direction of the original observation of the unstable nature of log gravity. Very importantly, the Ewens sampling formula is also well known to govern the distribution on partitions in random processes of fragmentation and coagulation \cite{gnedin2007poisson}. Furthermore, it is known that several models of random fragmentation involve structures such as random trees and forests \cite{bertoin2006random}. From these findings, our combinatorial construction in \cite{Mvondo-She:2022jnf} finds a natural interpretation as a tree representation of fragmentations. We argue that the unstable propagation of the soliton formed by the collection of fields at the single particle level $n=1$ gives rise to fragments of soliton clusters at the multiparticle levels $n \geq 2$. The soliton clusters are rooted trees partitioned into a collection of subtrees, and the coalescing (coagulating) subtrees are solitons pinned by point defects represented by the roots as the pinning sites. The unstable defect soliton clusters thus provide a disordered landscape whose random distribution is used in the next section to study the evolution of the fragmentation. 

In section 3, having established a connection between Hurwitz numbers and (Ewens) probability distribution, we use the Shannon information entropy as a measure of the evolution of the solitonic fragmentation in the theory. Since the inception of information theory, the Shannon information entropy \cite{shannon1948mathematical} has proved to be a considerably useful tool in quantifying the average information of a process in which outcomes have corresponding probabilities that add up to a distribution. At the junction between several fields \cite{lesne2014shannon}, it has been used in various forms. For instance, a particular realization of Shannon information entropy that provides an information measure of spatially localized physical configurations has been largely adopted in high energy physics under the name of configuration entropy, and used in connection with entropic measures of nonlinear scalar field models with spatially-localized energy solutions that include solitons and bounces in one spatial dimension and critical bubbles in three spatial dimensions \cite{Gleiser:2011di}, AdS–Schwarzschild black holes \cite{Braga:2016wzx}, the graviton Bose-Einstein condensate \cite{Casadio:2016aum}, the AdS/QCD correspondence \cite{Bernardini:2016hvx,Braga:2017fsb,Barbosa-Cendejas:2018mng,Bernardini:2018uuy,Ferreira:2019inu,Ferreira:2019nkz,Ferreira:2020iry,MarinhoRodrigues:2020ssq,MarinhoRodrigues:2020yzh,daRocha:2021imz,daRocha:2021ntm,Karapetyan:2021ufz,Karapetyan:2022rpl,daRocha:2022bnk}, or also in the analysis of Korteweg–de Vries solitons in the quark–gluon plasma \cite{GoncalvesdaSilva:2017bvk}. For applications of Shannon information entropy in other areas of science, including the analysis of heavy-ion collisions, see for instance \cite{Ma:2018wtw}. Here, we use Shannon information entropy as a measure of the disorder (or randomness) manifested in the generation of the unstable defect soliton clusters. Our information-theoretic formulation is based on describing the statistics of the fragmentation evolution by constructing a function that measures the difference between the axiomatic maximum entropy and the actual Shannon entropy at each level $n$. The motivation for deriving such a function is that the axiomatic maximum entropy value and the actual Shannon entropy at each level $n$ both increase at different rates. It therefore becomes useful to compare the gap between both values by means of difference. As the later is shown to increase with $n$, we find an interesting way of quantifying the decaying evolution of the unstable solitonic fragmentation.

In section 4, we take our probabilistic perspective of the log partition function a bit further, and discuss the appearance of a Bose-Einstein condensation in log gravity. Our motivation comes from Feynman's statistical model of collective excitations in systems of bosons, that originates from his study of the $\lambda$-transition in liquid helium \cite{Feynman:1953zza}. Feynman's formulation of Bose-Einstein condensation hinges on the idea of adapting his path-integral approach to quantum mechanics based on the quantum partition function into a statistical mechanics problem, resulting in the description of Bose systems in terms of large ensembles of cycles of various lengths, within which a random number of particles is spatially arranged according to all possible permutations of $n$-particles. In the context of log gravity, the interpretation is that at the critical point, the dynamics of soliton propagation is reflected by the emergence of fractions (fragments) of the $n$-particles as condensates, arranged in terms of Feynman graphs. The random permutations take place in a rooted tree configuration space, and a dynamical pinning mechanism arises at the root which represents the pinning site, leading to the formation of defect solitonic states. These solitons pinned by a point defect (the root) are the structures that realize the Bose-Einstein condensates.

In section 5, we finally summarize our results.

\section{The log partition function}
The 1-loop partition function of log gravity originally calculated with the form \cite{Gaberdiel:2010xv}

\begin{eqnarray}
\label{z tmg}
{Z_{\rm{CCTMG}}} (q, \bar{q})= \prod_{n=2}^{\infty} \frac{1}{|1-q^n|^2} \prod_{m=2}^{\infty} \prod_{\bar{m}=0}^{\infty} \frac{1}{1-q^m \bar{q}^{\bar{m}}}, \hspace{1cm} \rm{with} \hspace{0.25cm} q=e^{2i \pi \tau}. \bar{q}=e^{-2i \pi \bar{\tau}}.
\end{eqnarray}

\noindent From the identification of the first product as the three-dimensional gravity partition function $Z_{0,1}$ in \cite{Maloney:2007ud}, we have the convention 

\begin{eqnarray}
\label{z grav z log}
{Z_{\rm{CCTMG}}} (q, \bar{q})=  {Z_{\rm{gravity}}} (q, \bar{q}) \cdot {Z_{\rm{log}}} (q, \bar{q}),
\end{eqnarray}

\noindent with contributions 

\begin{eqnarray}
{Z_{\rm{gravity}}} (q, \bar{q})= \prod_{n=2}^{\infty} \frac{1}{|1-q^n|^2}, \hspace{0.5cm} \mbox{and}
\hspace{0.5cm} {Z_{\rm{log}}} (q, \bar{q}) = \prod_{m=2}^{\infty} \prod_{\bar{m}=0}^{\infty} \frac{1}{1-q^m \bar{q}^{\bar{m}}}.
\end{eqnarray}

\noindent The log contribution of the CCTMG partition function is redefined in terms of permutation cycles by considering the relation between permutations and integer partitions as follows. We first recall that if a positive integer $n$ is written as the sum of positive integers

\begin{eqnarray}
n =  \underbrace{1 + \cdots +1}_\text{$j_1$ times} + \underbrace{2 + \cdots +2}_\text{$j_2$ times} + \cdots + \underbrace{n}_\text{$j_n$ times},
\end{eqnarray}

\noindent then any partition in the set of all partitions $\mathcal{P}_n$ of $n$ is determined by a sequence $\left(j\right)_{k=1}^n$ of positive integers $j_k$ called occupation numbers or cycle counts of the partition which satisfy the constraint 

\begin{eqnarray}
\label{constraint}
\sum_{k=1}^n k j_k = n.    
\end{eqnarray}

\noindent Hence, a partition of a given integer $n$ is simply an expression of $n$ in terms of a sum of positive integers. 

\noindent A permutation with exactly $j_k$ cycles of length $k$ turns out to be of the same type as the sequence $\left(j\right)_{k=1}^n$ of cycle counts that determines a given partition. As a result, each partition corresponds to a conjugacy class of permutations whose number of elements is

\begin{eqnarray}
\label{N(n,j)}
 N(n,j) = \frac{n!}{\prod_{k=1}^n j_k! (k) ^{j_k}}.   
\end{eqnarray}

\noindent Under the rescaling of variables with the coordinate sequence $\left( \mathcal{G}_k \right)_{k=1}^n$ as 

\begin{eqnarray}
\label{G_n}
\mathcal{G}_k \left( q,\bar{q} \right) = \frac{1}{|1-q^k|^2}, 
\end{eqnarray}

\noindent the cycle decomposition of any permutation $\pi$ in the symmetric group $S_n$ of all permutations on integrers $1, \ldots , n$ is generated by the partition function according to the (formal) power series expansion

\begin{eqnarray}
 Z_{log} \left( \mathcal{G}_1, \ldots , \mathcal{G}_n  \right) = \exp \left(  \sum_{k= 1}^{\infty} \frac{\mathcal{G}_k \left( q^2 \right)^k}{k} \right) = 1 +  \sum_{n=1}^{\infty} \frac{1}{n!} \left( \sum_{\pi \in S_n} \prod_{k=1}^n \mathcal{G}_k^{j_k} \right)\left( q^2 \right)^n.  
\end{eqnarray}

\noindent The log partition function is furthermore expressible as a generating function of Hurwitz numbers. The Hurwitz enumeration problem, which arose in the theory of Riemann surfaces \cite{hurwitz1891riemann}, consists in counting the number of ways a given permutation can be written as a product of a minimal number of transpositions that generate the full symmetric group. The outcome of the counting is a numerical sequence known as Hurwitz numbers. Also expressible as a count of ramified coverings of Riemann surfaces, in our case, the Hurwitz numbers count disconected $n$-fold coverings of $\mathbb{CP}^1$ by itself. Starting from the multinomial coefficient $N(n,j)$, the log partition function reads 

\begin{subequations}
\begin{align}
Z_{log} \left( \mathcal{G}_1, \ldots , \mathcal{G}_n  \right) &= 1 + \sum_{n=1}^{\infty} \frac{1}{n!} \left( \sum_{\substack{\sum_{k=1}^n k j_k =n \\ n \geq 1\\j_k \geq 0 }} N(n,j) \prod_{k=1}^n \mathcal{G}_k^{j_k} \right) \left( q^2 \right)^n   \\
&= 1 + \sum_{n=1}^{\infty} \frac{1}{n!} \left( \sum_{\substack{\sum_{k=1}^n k j_k =n \\ n \geq 1\\j_k \geq 0 }} \frac{n!}{\prod_{k=1}^n j_k! (k) ^{j_k}} \prod_{k=1}^n \mathcal{G}_k^{j_k} \right) \left( q^2 \right)^n  
\end{align}
\end{subequations}

\begin{subequations}
\begin{align}
&= 1 + \sum_{n=1}^{\infty}  \left( \sum_{\substack{\sum_{k=1}^n k j_k =n \\ n \geq 1\\j_k \geq 0 }} \frac{1}{\prod_{k=1}^n j_k! (k) ^{j_k}} \prod_{k=1}^n \mathcal{G}_k^{j_k} \right) \left( q^2 \right)^n \\ 
&= 1 + \sum_{n=1}^{\infty}  \left( \sum_{\substack{\sum_{k=1}^n k j_k =n \\ n \geq 1\\j_k \geq 0 }} \left[ H^{\bullet}_{0 \xrightarrow[]{n} 0} \left( \left( [1]^{j_1}, [2]^{j_2}, \ldots \right),  \left( [1]^{j_1}, [2]^{j_2}, \ldots \right) \right) \right] \prod_{k=1}^n \mathcal{G}_k^{j_k} \right) \left( q^2 \right)^n. \label{Hurwitz tau function}
\end{align}
\end{subequations}

\noindent and under the constraint $\left( \ref{constraint} \right)$ the disconnected Hurwitz number expression takes the form

\begin{eqnarray}
\label{Hurwitz numbers}
H^{\bullet}_{0 \xrightarrow[]{n} 0} \left( \left( [1]^{j_1}, [2]^{j_2}, \ldots \right),  \left( [1]^{j_1}, [2]^{j_2}, \ldots \right) \right) = \prod_{k=1}^n \frac{1}{j_k!  (k)^{j_k}},
\end{eqnarray}

\noindent where the $\left( [1]^{j_1}, \ldots, [k]^{j_k} \right)$ associated to the monomials $\prod_{k=1}^n \mathcal{G}_k^{j_k}$ are such that $[k]^{j_k}= \overbrace{k, \ldots. k}^{j_k ~ \rm{times}}$.

\noindent The partition function can also be expressed in terms of rooted trees as

\begin{eqnarray}
\label{Hurwitz partition function}
Z_{log} \left( l_1, \ldots, l_n \right) =  1 + \sum_{n=1}^{\infty}  \left( \sum_{\substack{\sum_{k=1}^n k j_k =n \\ n \geq 1\\j_k \geq 0 }} \left[ H^{\bullet}_{0 \xrightarrow[]{n} 0} \left( \left( [1]^{j_1}, [2]^{j_2}, \ldots \right),  \left( [1]^{j_1}, [2]^{j_2}, \ldots \right) \right) \right] B_+ \left( \prod_{k=1}^n l_k^{j_k} \right) \right) \left( q^2 \right)^n,
\end{eqnarray}

\noindent in terms of the same disconnected Hurwitz numbers $H^{\bullet}_{0 \xrightarrow[]{n} 0} \left( \left( [1]^{j_1}, [2]^{j_2}, \ldots \right),  \left( [1]^{j_1}, [2]^{j_2}, \ldots \right) \right)$ but this time, the $\left( [1]^{j_1}, \ldots, [k]^{j_k} \right)$ are associated to the trees $B_+ \left( \prod_{k=1}^n l_k^{j_k} \right)$.

Let us look at an example and consider the level $n=4$. The integer $4$ can be written as 

\begin{subequations}
\begin{align}
1+1+1+1, \\ 1+1+2, \\ 1+3, \\ 2+2, \\ 4, 
\end{align}    
\end{subequations}

\noindent with as corresponding partitions the sequences 

\begin{subequations}
\begin{align}
(4,0,0,0), \\ (2,1,0,0), \\ (1,0,1,0), \\ (0,2,0,0), \\ (0,0,0,1), 
\end{align}    
\end{subequations}

\noindent respectively, and the Hurwitz numbers take the general form

\begin{eqnarray}
 \frac{1}{j_1! j_2! j_3! j_4! \left( 1 \right)^{j_1} \left( 2 \right)^{j_2} \left( 3 \right)^{j_3} \left( 4 \right)^{j_4}}.   
\end{eqnarray}

\noindent Table \ref{table 1} gives the Hurwitz numbers together with their associated $\mathcal{G}_k$-monomial products and rooted trees.

\begin{table}[h]
\begin{center}
\renewcommand*{\arraystretch}{1.9}
\begin{tabular}{|c|c|c|} 
 \hline
 Hurwitz numbers & $\mathcal{G}_k$-monomial products & Rooted trees  \\ [0.5ex] 
 \hline\hline
 $H^{\bullet}_{0 \xrightarrow[]{4} 0} \left( \left( 1,1,1,1 \right),  \left( 1,1,1,1 \right) \right) = \frac{1}{24}$ & $\mathcal{G}_1^4$ & 
 \begin{forest}
for tree={circle,draw,fill,minimum width=2pt,inner sep=0pt,parent anchor=center,child anchor=center,s sep+=0pt,l sep-=5pt,grow=north,}
[, for tree={l=0} [][][][]] 
\end{forest}
 \\ [1ex]
 \hline
 $H^{\bullet}_{0 \xrightarrow[]{4} 0} \left( \left( 1,1,2 \right),  \left( 1,1,2 \right) \right)= \frac{1}{4}$ & $\mathcal{G}_1^2 \mathcal{G}_2$ & 
 \begin{forest}
for tree={circle,draw,fill,minimum width=2pt,inner sep=0pt,parent anchor=center,child anchor=center,s sep+=0pt,l sep-=5pt,grow=north,}
[, for tree={l=0} [[]] [] []] 
\end{forest}
 \\ [1ex]
 \hline
 $H^{\bullet}_{0 \xrightarrow[]{4} 0} \left( \left( 1,3 \right),  \left( 1,3 \right) \right)= \frac{1}{3}$ & $\mathcal{G}_1 \mathcal{G}_3$ & 
 \begin{forest}
for tree={circle,draw,fill,minimum width=2pt,inner sep=0pt,parent anchor=center,child anchor=center,s sep+=0pt,l sep-=5pt,grow=north,}
[, for tree={l=0} [[[]]] []] 
\end{forest}
 \\ [1ex]
 \hline
 $H^{\bullet}_{0 \xrightarrow[]{4} 0} \left( \left( 2,2 \right),  \left( 2,2 \right) \right)= \frac{1}{8}$ & $\mathcal{G}_2^2 \mathcal{G}_2^2$ & 
 \begin{forest}
for tree={circle,draw,fill,minimum width=2pt,inner sep=0pt,parent anchor=center,child anchor=center,s sep+=0pt,l sep-=5pt,grow=north,}
[, for tree={l=0} [[]] [[]]] 
\end{forest}
 \\ [1ex] 
 \hline
 $H^{\bullet}_{0 \xrightarrow[]{4} 0} \left( \left( 4 \right),  \left( 4 \right) \right)= \frac{1}{4}$ & $\mathcal{G}_4$ & 
 \begin{forest}
for tree={circle,draw,fill,minimum width=2pt,inner sep=0pt,parent anchor=center,child anchor=center,s sep+=0pt,l sep-=5pt,grow=north,}
[, for tree={l=0} [[[[]]]] ] 
\end{forest}
 \\ [1ex] 
 \hline
\end{tabular}
\caption{\label{table 1} $n=4$ Hurwitz numbers, associated $\mathcal{G}_k$-monomial products and rooted trees.}
\end{center}
\end{table}

We would like at this stage to point out an interesting probabilistic interpretation of the above results. Via the multinomial coefficient $N(n,j)$, the log partition function actually shows an equivalence between random mappings specified by disconnected $n$-coverings and expansions over rooted trees. It turns out that these random combinatorial objects are built around probability distributions where Hurwitz numbers can play a central role. The probabilistic structure of these combinatorial objects associated to Hurwitz multinomial expansions have been addressed in Pitman's elegant work \cite{pitman1997asymptotic, pitman1997abel,pitman2001random,pitman2002forest}, as well as in \cite{okounkov2001gromov} in connection with integrable hierarchies. Related to our work, we highlight a remarkable connection between the Hurwits numbers and the Ewens sampling formula. In 1972, Ewens \cite{ewens1972sampling} derived a probability distribution for vectors of the type $j= \left( j_k \right)_{k=1}^{n}$, where $j_k$ counts the number of selectively neutral alleles represented $k$ times in a sample of $n$ genes taken from a large population. Equivalently, the Ewens sampling formula can be reformulated as the distribution of the cycle counts of a permutation $\pi \in S_n$, decomposed as a product of
cycles. According to \cite{arratia2003logarithmic}, for the uniform and random choice of a given permutation $\pi \in S_n$, the distribution of the cycle counts is given by

\begin{eqnarray}
\label{ESF}
\mathbb{P} \left[ j = \left( j_k \right)_{k=1}^{n}  \right]  = \mathbbm{1} \left\{ \sum_{k=1}^n k j_k =n  \right\} \prod_{k=1}^n \frac{1}{j_k!  (k)^{j_k}}
\end{eqnarray}

\noindent where the indicator $\mathbbm{1} \left\{ \cdot \right\}$ is defined as

\begin{eqnarray}
 \mathbbm{1} \left\{ A \right\} = 
 \left\{
    \begin{array}{ll}
        1 & \mbox{if $A$ is true,} \\
        0 & \mbox{otherwise.}
    \end{array}
\right.
\end{eqnarray}

\noindent The probability distribution $\left( \ref{ESF} \right)$ is referred to as the Ewens Sampling Formula with parameter $\theta =1$. The Ewens Sampling formula with parameter $\theta \geq 0$ reads 

\begin{eqnarray}
\mathbb{P} \left[ j = \left(j_1, \ldots , j_n  \right)  \right]  = \frac{n!}{\theta \left( \theta +1 \right) \cdots \left(  \theta + n -1 \right)}\mathbbm{1} \left\{ \sum_{k=1}^n k j_k =n  \right\} \prod_{k=1}^n \frac{\theta}{j_k!  (k)^{j_k}}.
\end{eqnarray}

The Ewens sampling formula is known to govern the distribution on partitions of $\left\{ 1, \ldots,n \right\}$ for certain partition-valued fragmentation processes \cite{mekjian1991cluster,gnedin2007poisson,pitman2006combinatorial}. It is also known that the models under which the Ewens sampling formula holds exhibit a diffusion process \cite{griffiths2005ewens}. 

We argue that the unstable propagation of the seed soliton composed of the collection of fields at the single particle level $n=1$ and represented by $\mathcal{G}_1$ in the log partition function, gives rise to a physical process of fragmentation, in which soliton clusters are produced at the multiparticle levels $n \geq 2$. The fragmented soliton clusters are realized as rooted trees partitioned into a collection of subtrees, and the juxtaposed subtrees are solitons pinned by point defects represented by the roots as the pinning sites. Ultimately, the unstable defect soliton clusters provide a disordered landscape whose distribution is used in the next section to study the evolution of the unstable fragmentation process.

\section{Shannon information entropy as a measure of the evolution of the unstable defect soliton clusters}

Shannon's logarithmic nature of the measure of information proceeds from his 1948 celebrated paper \cite{shannon1948mathematical}. As a statistical entropy, Shannon information entropy is a logarithmic measure of the average information content in a collection of probabilities constrained in a distribution. In other words, it is a measure of disorder, or chaoticity, or ignorance about a certain data, in the form of uncertainty. Due to its applied flavor, Shannon information entropy finds relevance in the theory of dynamical systems, and can be used to study the evolution of a quantity or of a process. In our case, we use Shannon information entropy to assess the evolution of the disorder or randomness arising in the generation of defect soliton clusters as $n$ grows.

Let 

\begin{eqnarray}
\mathbb{D}_n = \left\{ \left( p_1,p_2, \ldots , p_n \right) | p_k\geq 0 , 0 \leq k \leq n, \sum_{k=1}^{n} p_k=1  \right\}, \hspace{1cm} n \geq 1    
\end{eqnarray}

\noindent denote the set of $n$-component discrete probability distributions. The Shannon entropy $\mathcal{H}_n : \mathbb{D}_n \mapsto \mathbb{R}$ is defined by

\begin{eqnarray}
\mathcal{H}_n \left( p_1,p_2, \ldots , p_n \right) = - \sum _{k=1}^{n} p_k \log_2 p_k, \hspace{1cm} n \geq 1,   
\end{eqnarray}

\noindent with by convention $0 \log_2 0 = 0$. Two extreme scenarios are presented below. 

Suppose first, that 

\begin{eqnarray}
p_k = \left\{ \begin{array}{l}1 \hspace{0.5cm} {\rm{for}} \hspace{0.25cm} k=1, \\ 0 \hspace{0.5cm} {\rm{for}} \hspace{0.25cm} 2 \leq k \leq n. \end{array} \right.
\end{eqnarray}

\noindent In information-theoretic language, it means that there is complete certainty about the occurrence of an event. In such case, the Shannon entropy $\mathcal{H}_n \left( 1, 0,0,0, \ldots \right)$ is just zero.

On the other hand, suppose that 

\begin{eqnarray}
p_k = \frac{1}{n} \hspace{0.5cm} {\rm{for}} \hspace{0.25cm} 1 \leq k \leq n.
\end{eqnarray}

\noindent In this case, the events $1, \ldots , n$ have equal likelihood, and the entropy (uncertainty) $\mathcal{H} \left( p_1, \ldots, P_n \right) = \mathcal{H} \left( \frac{1}{n}, \ldots, \frac{1}{n} \right)$ is maximized over the probability vectors $p_1, \ldots, p_n$ of length $n$.

The above extreme cases arise at the $n=1$ (single) particle and at the $n=2$ multi particle levels, respectively. At the single particle level, there is complete certainty about the state of the system, and the entropy is null. At the $n=2$ multi particle level, the amount of uncertainty about the state of the system is equally distributed among the microstates of the 2-particle system, and the entropy has reached its largest possible value. The latter case features the connection between entropy and symmetry. Indeed, up to permutation of $p_1$ and $p_2$, the symmetry of the distribution is high since the probabilities are identical, and indicates the indistinguishability of the system that corresponds to a total loss of information, hence to a state of maximum entropy in the $n=2$ multi-particle system. 

Between the two aforementioned extreme cases, the multiparticle levels $n \geq 3$ have distributions for which not all probabilities $p_k$ are identical. This is an indication that the Shannon entropy is no longer maximum from $n \geq 3$. We would like to assess the change in entropy, in a way that tells us something about the evolution of the diffusive instability. A naive option would be to just calculate the Shannon entropy at each level $n$. In so doing, we immediately notice that the Shannon entropy of the distribution increases with $n$. However, this would not teach us much, as the maximum entropy also grows with $n$, but at a different rate than the entropy from the actual distribution in the system. To have a good indicator of the instability evolution, we rather derive a function that measures the difference between maximum and actual entropies at each level $n$. We derive it as follows. 

We first recall that given the set $\mathcal{P}_n$ of all partitions of $n$, the Ewens sampling Formula $\mathbb{P}$ with parameter $\theta =1$ is a probability distribution obtained by sampling uniformly over a partition $1^{j_1} 2^{j_2} \cdots n^{j_n}$ of $n$. This means that the space of probabilities $ p_k $ we need to consider is restricted by  

\begin{eqnarray}
1 \leq p_k \leq  \mathtt{p} (n),  
\end{eqnarray}

\noindent where $\mathtt{p} (n)$ is the number of partitions of $n$.

\noindent The expression of the maximum entropy at level $n$ becomes

\begin{eqnarray}
\mathcal{H}^{\rm{max}}_n \left[ \left( p_k \right)_{k=1}^{\mathtt{p} (n)} \right] = \mathcal{H} \left( \frac{1}{\mathtt{p} (n)}, \ldots, \frac{1}{\mathtt{p} (n)} \right) = \log_2 \left[ \mathtt{p} (n) \right].
\end{eqnarray}

\noindent The function $\mathcal{H}^{\Delta}_n \left[ \left( p_k \right)_{k \in \mathcal{P}_n} \right]$ which measure the difference between maximum and actual entropies at each level $n \geq 2$ takes the form

\begin{subequations}
\begin{align}
\mathcal{H}^{\Delta}_n \left[ \left( p_k \right)_{k \in \mathcal{P}_n} \right] &= \mathcal{H} \left( \frac{1}{\mathtt{p} (n)}, \ldots, \frac{1}{\mathtt{p} (n)} \right) -  \mathcal{H}_n \left[ \left( p_k \right)_{k=1}^{\mathtt{p} (n)} ~|~ p_k \in \mathbb{P} \right]  \\ 
&= \log_2 \left[ \mathtt{p} (n) \right] - \left( - \sum _{p_k \in \mathbb{P}} p_k \log_2 p_k  \right) \\
&= \log_2 \left[ \mathtt{p} (n) \right] +   \log_2 \left( \prod _{p_k \in \mathbb{P}} p_k^{p_k} \right) \\
&= \log_2 \left( \mathtt{p} (n) \cdot \prod _{p_k \in \mathbb{P}} p_k^{p_k} \right), \hspace{1cm} n \geq 2. 
\end{align}   
\end{subequations}

\noindent Table \ref{table 2} gives an example of the computations of $\mathcal{H}^{\Delta}_n \left[ \left( p_k \right)_{k \in \mathcal{P}_n} \right]$ for $n=2.3.4$.

\begin{table}[h]
\begin{center}
\renewcommand*{\arraystretch}{1.8}
\begin{tabular}{|c|c|c|c|} 
 \hline 
 n & $\mathcal{H}^{\rm{max}}_n \left[ \left( p_k \right)_{k=1}^{\mathtt{p} (n)} \right]$ & $\mathcal{H}_n \left[ \left( p_k \right)_{k=1}^{\mathtt{p} (n)} ~|~ p_k \in \mathbb{P} \right]$ & $\mathcal{H}^{\Delta}_n \left[ \left( p_k \right)_{k \in \mathcal{P}_n} \right]$ \\  
 \hline\hline 
 2 & $- \log_2 \left[ \left( \frac{1}{2} \right)^{\frac{1}{2}} \left( \frac{1}{2} \right)^{\frac{1}{2}} \right]= \log_2 2$  & $- \log_2 \left[ \left( \frac{1}{2} \right)^{\frac{1}{2}} \left( \frac{1}{2} \right)^{\frac{1}{2}} \right]$  & 0 \\ 
 \hline
 3 & $- \log_2 \left[ \left( \frac{1}{3} \right)^{\frac{1}{3}} \left( \frac{1}{3} \right)^{\frac{1}{3}} \left( \frac{1}{3} \right)^{\frac{1}{3}} \right]= \log_2 3$ & $- \log_2 \left[ \left( \frac{1}{6} \right)^{\frac{1}{6}} \left( \frac{1}{2} \right)^{\frac{1}{2}} \left( \frac{1}{3} \right)^{\frac{1}{3}} \right]$ & 0.1258145837 \\
 \hline
 4 & $- \log_2 \left[ \left( \frac{1}{5} \right)^{\frac{1}{5}} \left( \frac{1}{5} \right)^{\frac{1}{5}} \left( \frac{1}{5} \right)^{\frac{1}{5}} \left( \frac{1}{5} \right)^{\frac{1}{5}} \left( \frac{1}{5} \right)^{\frac{1}{5}} \right]= \log_2 5$ & $- \log_2 \left[ \left( \frac{1}{24} \right)^{\frac{1}{24}} \left( \frac{1}{4} \right)^{\frac{1}{4}} \left( \frac{1}{3} \right)^{\frac{1}{3}} \left( \frac{1}{8} \right)^{\frac{1}{8}} \left( \frac{1}{4} \right)^{\frac{1}{4}} \right]$ & 0.2275671569 \\ 
 \hline
\end{tabular}
\caption{\label{table 2} Values of $\mathcal{H}^{\rm{max}}_n \left[ \left( p_k \right)_{k=1}^{\mathtt{p} (n)} \right]$, $\mathcal{H}_n \left[ \left( p_k \right)_{k=1}^{\mathtt{p} (n)} ~|~ p_k \in \mathbb{P} \right]$, and $\mathcal{H}^{\Delta}_n \left[ \left( p_k \right)_{k \in \mathcal{P}_n} \right]$ for $n=2.3.4$.}
\end{center}
\end{table}

 \noindent The null value of $\mathcal{H}^{\Delta}_n \left[ \left( p_k \right)_{k \in \mathcal{P}_n} \right]$ at $n=2$ indicates that the entropy is maximum at that level. In other words, the disorder associated with the fragmentation of the seed soliton whose collective behavior is represented by $\mathcal{G}_1$ in the partition function into unstable defect soliton clusters represented by $\mathcal{G}_2$ and $\mathcal{G}_1^2$ is maximum. From $n \geq 3$, the positive and increasing values of the logarithmic function $\mathcal{H}^{\Delta}_n \left[ \left( p_k \right)_{k \in \mathcal{P}_n} \right]$ indicate that it monotonically increases with $n$. This shows that the Shannon entropy values with probabilities constrained by the Ewens distribution slowly move away from the maximum entropy values as $n$ increases, and that the defect soliton clusters generated slowly become less unstable.   

\section{Bose-Einstein condensation in configuration space}
The Bose-Einstein condensation (BEC) in an ideal gas of bosons was predicted in the middle of the 1920s based on ideas related to the statistical description of quanta of light \cite{Bose:1924mk,einstein1925quantum}, and experimentally observed for cold atomic gases in magnetic traps about seventy years later \cite{Anderson:1995gf,Bradley:1995zz,Davis:1995pg,Dalfovo:1999zz}. The main idea of the Bose-Einstein condensation is that if the density of particles exceeds a certain critical value, a fraction (fragment) of the whole amount of particles clusters (condenses) in the lowest eigenstate. The approach we consider to discuss Bose-Einstein condensation in log gravity is based on a probabilistic interpretation of the phase transition which goes back to Feynman's work on $\lambda$-transition in liquid helium, and his proposal to write the partition function as a path integral over trajec-
tories of the helium atoms, showing that a Bose-Einstein phase transition appears because of the symmetry statistics of atoms that can be permuted with each other. 

The historical starting point is Feynman's formulation of a path-integral treatment of quantum mechanics within which transition probability amplitudes can be computed in terms of superpositions of classical paths, accounting for all possible trajectories that contribute to the quantum process through constructive superposition \cite{Feynman:1948ur}. This probabilistic formulation of a quantum observable can be extended to statistical mechanics by mapping quantum particles in a many-body system to paths in space and imaginary-time, in such a way that the quantum system is described in terms of classical trajectory configurations.

Shortly after his development of the path integral formalism, Feynman adapted it to statistical mechanics by introducing a concept of permutation cycles in the partition function of a boson system. The key point underpinning Feynman's approach is that, considering the ground state wavefunction of liquid Helium as non-degenerate, real positive and without nodes. and taking into account that the interatomic interaction of Helium are repulsive in nature, the wavefunction vanishes within an interatomic range \cite{feynman1955chapter}. Because the Helium atom has spin zero, the wavefunction is totally symmetric with respect to the positions of the atoms. From there, Feynman argued that at low energy, the excitations are only composed of density waves (phonons) with a linear dispersion law, and that as the Bose-Einstein statistics would not allow long wavelength (low energy) excitations, any long distance movement of Helium atoms consists of a permutation of the atoms that leaves the wavefunction unchanged. 

Of geometrical importance in Feynman's approach is the configuration space of physical fields, which is a gauge invariant space with the symmetric group $S_n$ as a gauge group. In a configuration of particles characterized by the particles' coordinates modulo the permutation symmetry, the relationship between the Bose-Einstein condensation and the partition function of the ideal Bose gas in the $\lambda$-transition can be established by considering the symmetrized states of a system with $n$ identical bosons. The permutation symmetry in the many-particle configurations is extracted from the partition function $Z_n$ by writing it in terms of a cycle structure where the permutation $\pi \in j_1, j_2, \ldots , j_n$ contains $j_1$ 1-cycles, $j_2$ 2-cycles, etc ... Then, given the configuration set $\mathcal{C} \left( j_1, j_2, \ldots  \right)$ of any set of non-negative integers satisfying the constraint $\left( \ref{constraint} \right)$, the partition function becomes

\begin{eqnarray}
Z_n = \frac{1}{n!}   \sum_{\left\{  j_1, j_2, \ldots \right\}} \mathcal{C} \left( j_1, j_2, \ldots  \right) \prod_{k} Z \left( \pi \right)^{j_k}. 
\end{eqnarray}

\noindent Feynman \cite{feynman2018statistical} and Matsubara \cite{matsubara1951quantum} showed that the quantity $\mathcal{C} \left( j_1, j_2, \ldots  \right)$ is identical to $N(n,j)$ in Eq. $\left( \ref{N(n,j)}\right) $. This establishes a connection betweeen our interpretation of the collective excitations in the log sector as a grand-canonical ensemble of permutation trees, and Feynman's argument that Bose-Einstein condensation is understood as the occurrence of large cycles of bosons permuted in imaginary time, to infer the realization of Bose-einstein condensates as fragmentation trees in log gravity. 

\section{Summary and outlook}
A probabilistic stance of the log partition function of CCTMG was given in this work, and is summarized below. 

\begin{itemize}
\item Previously studied as topological invariants, the Hurwitz numbers reappear in this paper as probabilities in a multinomial distribution that governs the permutation cycles present in the rooted tree configuration space described in our earlier works. 

\item The multinomial distribution was shown to be directly related to the Ewens sampling formula with parameter $\theta = 1$, also called the Multivariate Ewens distribution in Bayesian statistics, which beyond its extensive use in mathematical biology and statistics, generally plays a major role in the theory of partition structures and their applications in processes of fragmentation and coagulation, and in random trees. The appearance of the Ewens distribution in the theory, as well as the description of the partition  function in terms of the (potential) Burgers hierarchy clearly indicate a diffusive behavior that maps out the instability originally observed in \cite{Grumiller:2008qz}. 

\item The statistical formulation of Burgers turbulence and its relation to (elastic) manifolds pinned by quenched disorder (see for instance \cite{wiese2022theory} for a recent review) have previously been studied from the perspective of directed polymers \cite{mezard1998disordered}, disordered trees and traveling waves \cite{derrida1988polymers}, contexts within which the quenched disorder appears to be provided by the pinning of a fraction of the particles. We infer that the original unstable aspect of the theory due to the appearance of the logarithmic mode at the critical point corresponds to the instability of the seed ($n=1$) soliton as collective fields. The unstable (diffusive) propagation of the seed soliton engenders soliton clusters that can be arranged as random (fragments of) rooted trees characterizing the disordered landscape in the theory. The rooted tree configuration space is realized by the pinning of randomly selected point-like and line-like solitonic particles at the root which constitutes the pinning site. 

\item Given the probabilistic distribution it falls under, as $n$ grows, the evolution of the process influenced by the quenched disorder, $i.e$ the random pinning was studied using Shannon information entropy. At $n=1$, the probability of formation of the log mode together with its descendants is $100 \%$, and its probability is $p_1 = 1$. In informtion-theoretic parlance, for such an event, there is zero surprise, or zero information, and the entropy of formation of the collective fields at $n=1$ as the seed soliton is therefore zero. From $n=2$, there is a growing range of possibilities with particular probabilities. With these possibilities enters the notion of symmetry, related to the similarity of the probabilities. At $n=2$, the two possible outcomes having identical probabilities indicates that the disorder is maximum. From $n \geq 3$, the greater the number of possible outcomes, the greater the value of the entropy, however, the actual values of the entropy at each $n$-level tend to grow away from the corresponding $n$-level maximum entropy values. This analysis gives a perspective of the evolution towards stability in the generation of the defect soliton clusters.  

\item The model of random permutations in configuration space originating from Feynman's path-integral approach to study bosonic systems collective excitations was used to infer the presence of Bose-Einstein condensates in log gravity, from the occurrence of permutation cycles in the log partition function.  

\item Away-from-equilibrium macroscopic phenomena naturally occur in various branches of basic and applied sciences, including fluid dynamics. Burgers turbulence appeared as one of the simplest instances of a nonlinear system out of equilibrium, and describes a host of seemingly unrelated phenomena. Indeed, apart from modeling turbulence in fluid dynamics, it has also been used to describe large scale pattern formation in cosmology, vortex lines in random media, and growth phenomena \cite{bec2007burgers,bonkile2018systematic}. The results obtained in this work indicates an out-of-equilibrium dynamics which corresponds to a type of shock-wave damping of the instability in a diffusive process.

\end{itemize}

Finally, from this work, a question arises: usually, in theories where an unstable configuration occurs, the later runs towards some (meta-)stable configuration. It has however been unclear from the earlier literature
on this subject, where the unstable theory is supposed to run to. We hope our work can provide further insight on this yet elusive point in future.

\paragraph{Acknowledgements} The author would like to thank the anonymous referee for useful suggestions. The author is grateful for the time spent at the 12th Joburg Workshop on String Theory, Gravity and Cosmology where this work started, and at the 5th Mandelstam Theoretical Physics School and Workshop, where it approached its conclusion. This work is supported by the South African Research Chairs initiative of the Department of Science and Technology and the National Research Foundation. The support of the DSI-NRF Centre of Excellence in Mathematical and Statistical Sciences (CoE-MaSS) towards this research is hereby acknowledged. Opinions expressed and conclusions arrived at, are those of the author and are not necessarily to be attributed to the CoE.

\clearpage

\bibliographystyle{utphys}
\bibliography{sample}
\end{document}